\documentclass[letterpaper, 10 pt, conference]{ieeeconf}

\IEEEoverridecommandlockouts                              

\usepackage{url}
\usepackage{tikz}
\usepackage{float}
\usepackage{multirow}

\newcommand{\prefig}[0]{}
\newcommand{\postfig}[0]{}

\title{\LARGE \bf
Detection of collaborative activity with Kinect depth cameras}

\author{Lo\"ic~Sevrin,
        Norbert~Noury, 
        Nacer~Abouchi, 
        Fabrice~Jumel,
        Bertrand~Massot,
        and~Jacques~Saraydaryan
\thanks{L. Sevrin is with the University of Lyon, INL, UCBL
{(e-mail: loic.sevrin@univ-lyon1.fr)}.}
\thanks{N. Noury is with the University of Lyon, INL, UCBL
{(e-mail: norbert.noury@univ-lyon1.fr)}.}
\thanks{N. Abouchi is with the University of Lyon, INL, CPE Lyon
{(e-mail: abouchi@cpe.fr)}.}
\thanks{F. Jumel is with the University of Lyon, CITI, CPE Lyon
{(e-mail: fabrice.jumel@cpe.fr)}.}
\thanks{B. Massot is with the University of Lyon, INL, INSA Lyon
{(e-mail: bertrand.massot@insa-lyon.fr)}.}
\thanks{J. Saraydaryan is with the University of Lyon, CITI, CPE Lyon
{(e-mail: jacques.saraydaryan@cpe.fr)}.}}

\begin{document}

\maketitle
\thispagestyle{empty}
\pagestyle{empty}


\definecolor{electricblue}{rgb}{0.49, 0.98, 1.0}

\definecolor{SkyBlue}{rgb}{0.53, 0.81, 0.92}

\definecolor{applegreen}{rgb}{0.55, 0.71, 0.0}

\definecolor{c1a}{HTML}{0099CC}
\definecolor{c1b}{HTML}{DBFFDB}
\definecolor{c1c}{HTML}{D4F1FF}
\definecolor{c1d}{HTML}{003399}

\definecolor{blue1}{HTML}{3498db}
\definecolor{blue2}{HTML}{2980b9} 
\definecolor{green2}{HTML}{27ae60}
\definecolor{black1}{HTML}{34495e}
\definecolor{red1}{HTML}{e74c3c}
\definecolor{turquoise2}{HTML}{16a085}


\begin{abstract}
The health status of elderly subjects is highly correlated to their activities together with their social interactions. Thus, the long term monitoring in home of their health status, shall also address the analysis of collaborative activities. This paper proposes a preliminary approach of such a system which can detect the simultaneous presence of several subjects in a common area using Kinect depth cameras. Most areas in home being dedicated to specific tasks, the localization enables the classification of tasks, whether collaborative or not. A scenario of a 24 hours day shrunk into 24 minutes was used to validate our approach. It pointed out the need of artifacts removal to reach high specificity and good sensitivity. 
\end{abstract}

\section{Introduction}
\label{part:introduction}

The improvement of the quality of life and of the healthcare system during the last decades enabled longer life.
In Europe, life expectancy at birth grew from 72 years in 1990 to 76 years in 2013 \cite{isbn9789241564885}.
Since people live longer, they are more affected by one or more chronic health conditions such as diabetes or Alzheimer's disease.
These diseases cannot be properly handled by hospitals since the latest were created for crisis management whereas the patients need long term health monitoring and care.

From this observation, researchers and physicians understood the need to develop healthcare management system at home. One of the key issues which must be addressed by this system is the daily health monitoring.
This monitoring has been attempted via several means which can mainly be classified into two parts:
The first option is to monitor physiological parameters of people like the brain temperature \cite{gehin:inserm-01154969} or the cardiac activity \cite{massot:hal-01226432}.
The second option is to monitor the activity of people like in the fusion of multiple sensors sources in a smart home to detect scenarios of activities \cite{noury2012fusion}.
This paper describes in part \ref{part:material-methods} the architecture of the LivINLab at the INL lab (Lyon, FR) where the activity can be monitored for several people at home at the same time.

In order to be accepted by the patients, this system must be easily understood. Hence, the affordance of the LivINLab was optimized in order to make it behave as a robot exoskeleton.
This robot is bio-inspired with afferent and efferent pathways sensing the environment and acting according to rules. 
It also needs to provide an abstraction layer for the interconnections of several protocols and devices.
This abstraction layer is based on the Robot Operating System (ROS) \cite{website:ros}. This open source firmware is widely spread in robotic systems, is highly compatible with many sensors and actuators, and provides several data processing tools.
The large community of developers behind ROS as well as its openness makes it a sustainable choice.
 
This concept of bio-inspired robot has already been described and compared to state of the art systems \cite{healthcom15submitted}.
Similar systems providing high level interconnections between several protocols exist.
The MPIGate Middleware is also based on ROS and creates a bridge between many kinds of communication protocols such as Bluetooth or ZigBee and ROS \cite{6331968}.
SYLPH is a non-standard platform managing Ambient Intelligence (AmI) \cite{5291723}. It provides similar features, but the use of customized, self-maintained tools makes it a risky choice in the long term, especially compared to ROS which benefits of a large developers community. The CAC-framework uses Node.js for gait analysis with Kinects \cite{7367714}. This approach is more sustainable than a full customized platform, but needs software development to access the Kinect (or any other sensor or actuator) data, whereas this part is directly available in ROS for many robots and sensors.

Most activity monitoring systems are designed to monitor only one person. Some of these systems would be able to monitor several people at the same time but the cooperation analysis is not a core feature.
As an example, in 2011, Noury et Al. monitored the electrical activity of a flat inhabited by one subject using only one sensor on the electric meter \cite{5742702}. They created an ambulatogram which showed the activity of the person at home along the days and highlighted the existence of circadian cycles in the daily activity, but only for one person at a time.

This paper proposes in part \ref{part:experimentation} a first approach of a cooperation analysis with the same kind of ambulatogram. To do so, a data fusion is performed between several people's ambulatogram.
This experiment uses Kinects depth cameras available in the LivINLab which enable the detection of the simultaneous presence of people in the same room. The scenario repeated 3 times by several people is based on a shrunk day where 24 hours are simulated in 24 minutes.

\section{Materials and Methods}
\label{part:material-methods}

\subsection{The LivINLab}

\begin{figure}
\prefig
\centering
\caption{The LivINLab logo}
\label{fig:livinlab-logo}
\includegraphics[width=4cm]{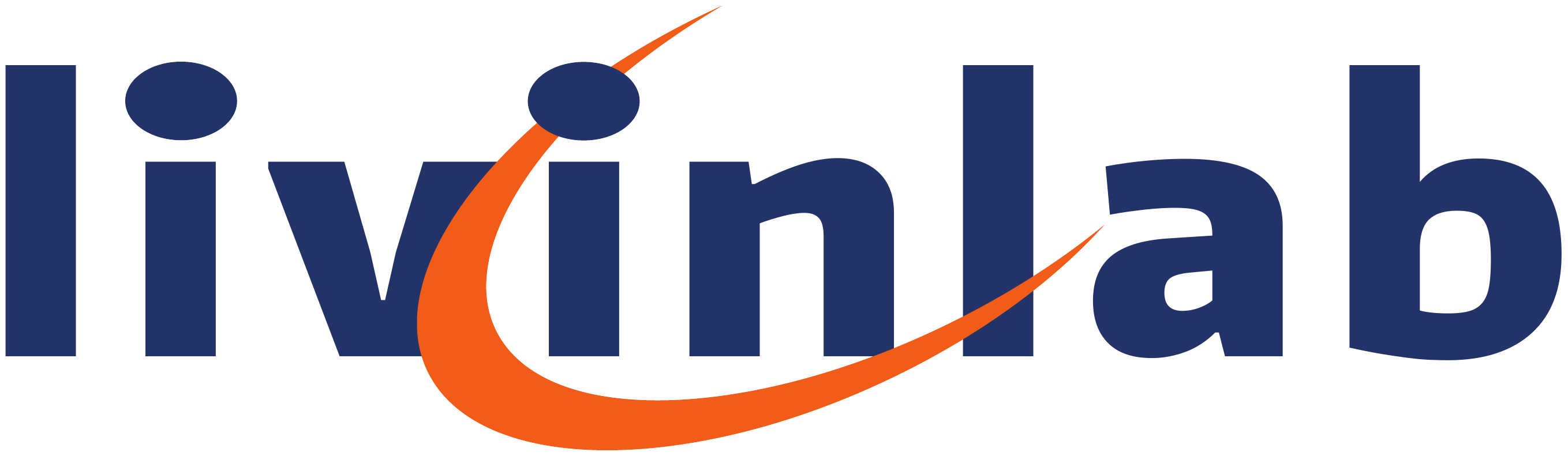}
\postfig
\end{figure}

The LivINLab (Fig. \ref{fig:livinlab-logo}) 
is a living lab for mobile health technologies, designed to be a place for quick prototyping, co-creation and experimentation.
It must free the working teams from most technological constraints.
Hence, it must be able to handle heterogeneous sensors and actuators in order to adapt easily to new technologies, protocols, sensors, ideas, etc.

The LivINLab is an apartment of $80 m^2$ composed of three rooms: the main room, the bedroom, and the bathroom. The main room includes the kitchen, the dining room, and the living room which are not delimited by walls (Fig. \ref{fig:livinlab-logo}).

\begin{figure}
\caption{The main room with a Kinect}
\label{fig:main-room}
\includegraphics[width=8.5cm]{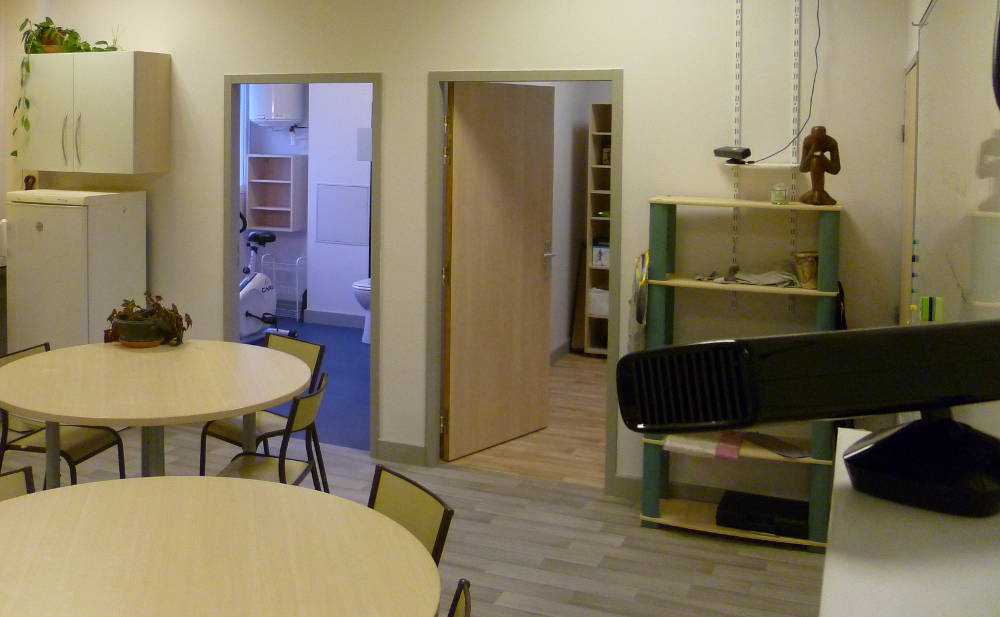}
\end{figure}

It includes many sensors such as three Kinect depth cameras for indoor positioning (two in the main room and one in the bedroom), and temperature, humidity and luminance sensors for environmental context sensing (Fig. \ref{fig:main-room}).

In this context of heterogeneous protocols and sensors, the architecture of the LivINLab was designed around the Robot Operating System (ROS). As discussed in part \ref{part:introduction}, this firmware benefits a large community of developers which made it able to communicate with many sensors and actuators, and often to process the data via available libraries. ROS is supported by several companies including Bosh and Qualcomm.

\subsection{ROS communication protocol}

\begin{figure}[]
\centering
\caption{ROS communication protocol, based on topics}
\label{fig:topics}
\begin{tikzpicture}[xscale=0.5, yscale=0.7]
\tikzstyle{nodeBase}=[draw, rounded corners=3pt];
\tikzstyle{nodeP}=[blue2, nodeBase];
\tikzstyle{nodeS}=[turquoise2, nodeBase];
\tikzstyle{nodeT}=[black, nodeBase];
\tikzstyle{legend}=[midway, below, text width=3cm, align=center];
\node (T) at (0,0) {\includegraphics[width=1.6cm]{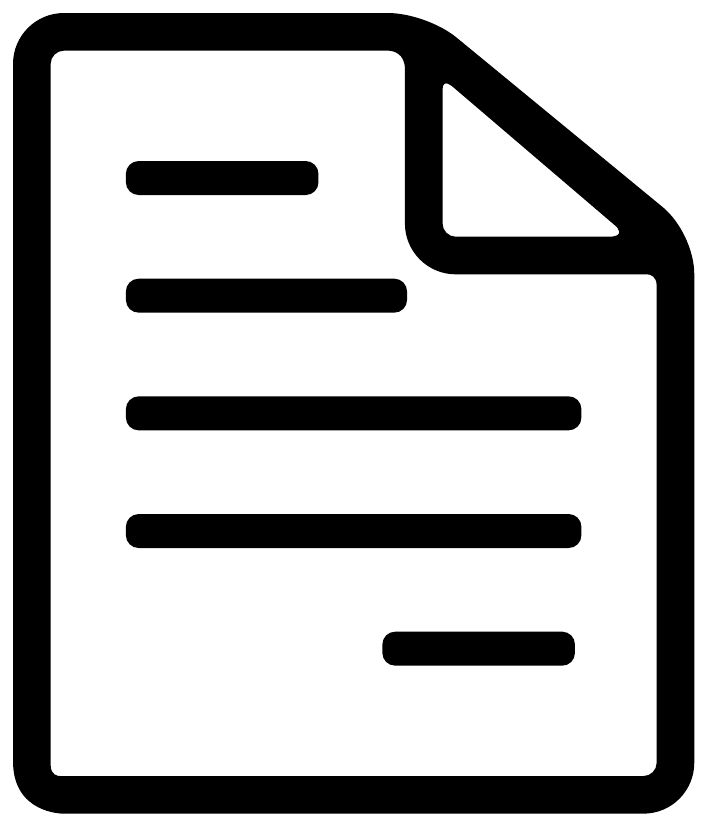}};
\node[nodeS] (S1) at (5,1) [draw,rounded corners=3pt, thick, fill=white]{subscriber node};
\node[nodeS] (S2) at (5,-1) [draw,rounded corners=3pt, thick, fill=white]{subscriber node};
\node[nodeP] (P) at (-5,0) [draw,rounded corners=3pt, thick, fill=white]{publisher node};
\draw[->, thick, nodeS] (T.east) to[bend right=10] (S1);
\draw[->, thick, nodeS] (T.east) to[bend left=10] (S2);
\draw[->, thick, nodeP] (P) to (T.west);

\draw[|-|, nodeP] (-7.5,-2) -- ++(5,0) node[legend]{write data};
\draw[|-|, nodeT] (-2,-2) -- ++(4,0) node[legend]{topic\\(abstraction layer)};
\draw[|-|, nodeS] (2.5,-2) -- ++(5,0) node[legend]{read data};


\end{tikzpicture}
\end{figure}
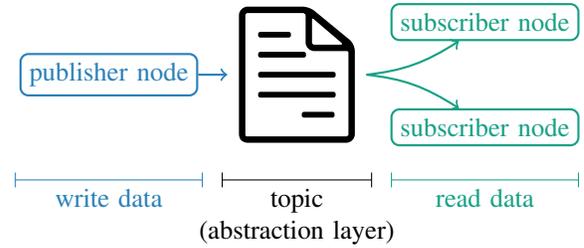

A ROS network is organized as follows: a ROS core running on a central computer manages several pieces of software (called ROS nodes) which can run independently on any computer of the local network. These nodes can publish data on topics which can be read by all the other nodes (Fig. \ref{fig:topics}). For a node reading a topic, the origin of the data does not matter. Only the data itself is used. Hence, every topic can be used as abstraction layers since changing the publisher will not affect the rest of the processing.

\subsection{Indoor Positionning with Depth Cameras}

A first step in the design of the living lab was the integration of several Kinects depth cameras \cite{Sevrin2015361}. It highlighted the relevance of ROS for interconnecting several sensors.
ROS nodes were already available for free to collect the video and depth streams, locate up to six people per Kinect and track their skeleton \cite{website:simple-openni}. 
Each Kinect is connected to a computer running these ROS nodes which publish the people positions on a ROS topic, over the Wi-Fi network. The development and maintenance costs were minimal, and speeded up the implementation.

The position of the subject's center of mass is calculated in the Kinect's landmark. Hence, every 3D position must be projected into the landmark of the apartment. For this purpose, ROS provides another efficient tool called tf \cite{6556373}. The latest enables the definition of geometrical transform (translation and rotation) from one landmark to another, in 3D. Then it can calculate any combinations of these transforms at any requested time.

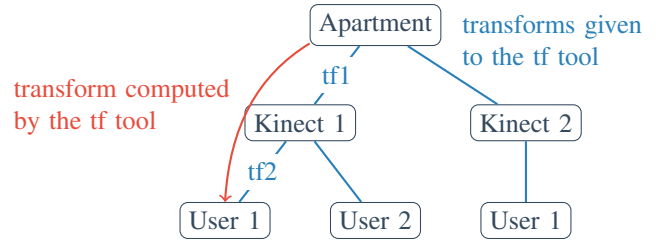
\begin{figure}
\prefig
\caption{The tf tree}
\label{fig:tftree}
\begin{center}
\begin{tikzpicture}[yscale=1.3]
\tikzstyle{node1}=[black1, draw,rounded corners=3pt];
\tikzstyle{link1}=[blue2, thick];
\tikzstyle{link2}=[->, red1, thick];
\node[node1] (A) at (1,0) {Apartment};
\node[node1] (K1) at (0,-1) {Kinect 1};
\node[node1] (U11) at (-1,-2) {User 1};
\node[node1] (U12) at (1,-2) {User 2};
\node[node1] (K2) at (3,-1) {Kinect 2};
\node[node1] (U21) at (3,-2) {User 1};
\draw[link1] (A) -- node[midway, fill=white]{tf1} (K1) -- node[midway, fill=white]{tf2} (U11);
\draw[link1](K1) -- (U12);
\draw[link1] (A) -- (K2) node[midway, anchor=south west, text width=5cm]{transforms given\\to the tf tool} to (U21);
\draw[link2] (A) to[bend right] node[midway, left, text width=3cm] {transform computed by the tf tool} (U11.north);
\end{tikzpicture}
\end{center}
\postfig
\end{figure}

For example, considering two transforms (Fig. \ref{fig:tftree}):
\begin{itemize}
\item The geometrical transform \textit{tf1} from the center of the apartment to the Kinect 1 (which is constant and can be measured),
\item The geometrical transform \textit{tf2} from the Kinect 1 to the center of mass of a person in the apartment (which is computed by the software positioning people in the Kinect's landmark),
\end{itemize}
Thanks to the tf tool, the positioning of the center of mass of the person in the landmark of the apartment is directly available through the composition of \textit{tf1} and \textit{tf2} .

If the requested transform is composed of several dynamic transforms (for example between two detected people), then the sample times when the transforms are recorded may not be the same. In this case, the tf tool will operate a linear interpolation on the samples in order to return an accurate result.

\section{Experimentation and Validation}
\label{part:experimentation}

\subsection{Collaborative Task Detection}

The monitoring of activity at home should not be reduced to the analysis of one person. Indeed, even for an elderly person living alone, his interactions with others (friends, nurses, etc.) are critical since the loss in social interactions is a key indicator of the degradation of the autonomy and health.
Hence, the ability of the LivINLab to 
individualize the simultaneous activities of several subjects in the same room
was chosen as a core feature. 
This enables the detection of similar tasks executed by several people, which could lead to collaboration analysis for a better granularity on health monitoring.

A subject activity is highly tied to where he is located. Indeed we cook in the kitchen, sleep in the bedroom, and take a shower in the bathroom. Hence the activity monitoring can be performed from locating the subjects inside the apartment. 
Hence, the analysis of simultaneous presence in the same area is a first indicator for the study of collaborative activity.

\subsection{Creation of Purpose Specific Areas}

Since the LivINLab 
does not include physical delimitations (walls) between some rooms such as the kitchen and the dinning room,
each area must be defined independently from the rooms themselves. 
The three Kinect depth cameras placed in the living lab cover four areas: the kitchen, the dining-room, the bedroom, and the office. 
The bathroom and the living-room are not covered.

\subsection{Scenario}

The experimentation is a scenario of 24 hours shrunk into 24 minutes. This kind of scenario is a model which highlights several transitions in a short time and improves the repeatability for easier preliminary analysis of the performances of the system \cite{6379452}.

The scenario describes a day of a person living alone and receiving a visitor during the afternoon. 
It was performed 3 times. 
It starts at 01:00 in order to be in a static phase when the subject is sleeping. In 24 minutes, several actions are performed including:
\begin{itemize}
\item Sleeping
\item Going to the toilets
\item Taking a shower
\item Going to the supermarket
\item Cooking
\item Eating
\item Watching TV
\item Chatting with a friend
\item etc...
\end{itemize} 

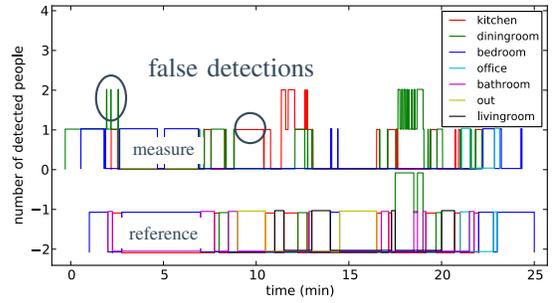
\begin{figure}
\centering
\caption{The raw ambulatogram (up) and the reference (down)}
\label{fig:raw-ambulatogram}

\begin{tikzpicture}
\tikzstyle{line1}=[black1, thick];
\tikzstyle{node1}=[line1, font=\scriptsize, fill=white];

\node at (0,0) {\includegraphics[width=8.5cm]{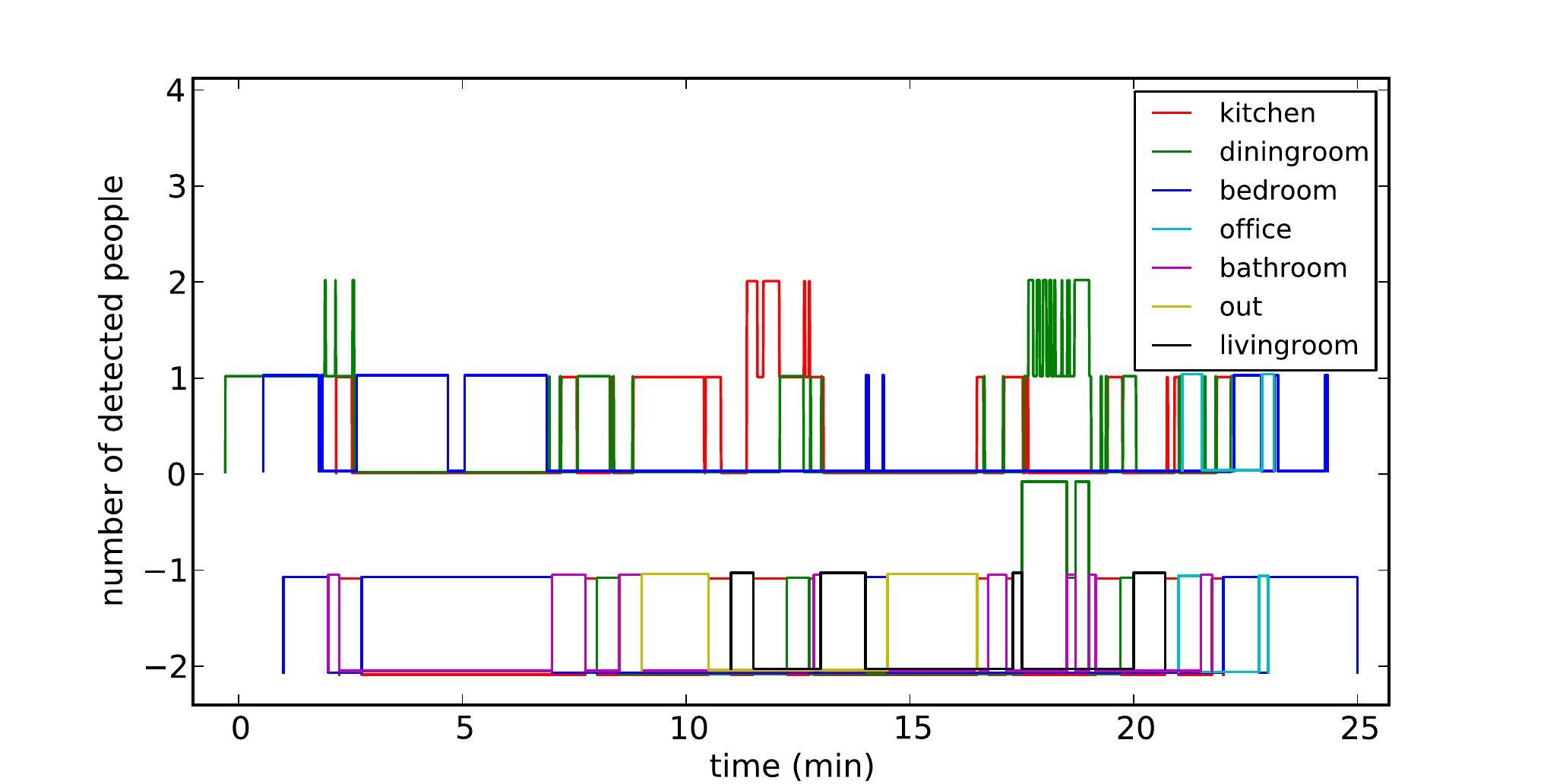}};
\draw[line1] (-2.4,0.5) ellipse (0.2cm and 0.3cm);
\draw[line1] (-0.55,0.1) ellipse (0.2cm and 0.2cm);
\node[line1] at (-0.8,0.9) {false detections};
\node[node1] at (-1.7,-0.2) {measure};
\node[node1] at (-1.7,-1.3) {reference};

\end{tikzpicture}
\end{figure}

The ambulatogram obtained shows the number of people detected per area along the day. 
The reference ambulatogram based on the scenario is also displayed (Fig. \ref{fig:raw-ambulatogram}).

The correlation between the reference and the measure is clear, even with several false detections (false positive) or missing ones (false negative). Since the living-room, the bathroom and the outside are not covered by the Kinects, those areas are only displayed in the reference. 

The detections of simultaneous presence in the same area can be seen. 
This results confirm the ability of our system to detect such an event, which is an indicator of simultaneous similar task execution. 
But as displayed in the reference, only the activity in the dining-room around 18:00 is collaborative. Hence the removal of some detection artifacts would improve the specificity of the detections, whether the activities are collaborative or not.

\subsection{Artifact Removal}

The false detection are caused by the algorithm processing the depth images of the Kinect. Two main sources of false detections were identified.

\begin{figure}
\centering
\caption{Choice of the perimeter as an indicator of false detection}
\label{fig:area-vs-perimeter}
\begin{tikzpicture}
\tikzstyle{node1}=[blue2];
\tikzstyle{node2}=[black1];
\tikzstyle{node3}=[node2, anchor=east];
\tikzstyle{line1}=[red1];

\begin{scope}[yscale=1, xshift=-3cm]
\draw[node3] (-3,-1.25) -- ++(6.5,0);
\node[node3] at (0,-1.5) {likely human};
\node[node3] at (0,-2) {high area};
\node[node3] at (0,-2.5) {high perimeter};
\end{scope}

\begin{scope}[xshift=-2cm]
\foreach \Point in {(0.1,0.1), (0,0), (0,0.2), (0.15,0.25), (0,0.25)}{
    \node[node1] at \Point {\textbullet};
}
\draw[node1] (0.1,0.1) -- (0,0) -- (0,0.2) -- (0.15,0.25) -- (0,0.25);
\draw[line1] (-0.1,-0.1) -- (-0.1,0.35) -- (0.30,0.35) -- (0.2,0.1) -- (0.1,-0.1) -- (-0.1,-0.1);
\draw[<-,  thick, node1] (-0.07, 0.2) -- ++(-0.5,0)  node[anchor=east, text width=4cm, align=right]{a measured position of the center of mass of a body};  
\node[node2] at (0,-1.5) {-};
\node[node2] at (0,-2) {-};
\node[node2] at (0,-2.5) {-};
\end{scope}

\begin{scope}[xshift=-1cm, yscale=1, yshift=-1cm]
\foreach \Point in {(0,0), (0,0.3), (0,0.75), (0.1,1.05), (0.15,1.35), (0.15,1.65)}{
    \node[node1] at \Point {\textbullet};
}
\draw[node1] (0,0) -- (0,0.3) -- (0,0.75) -- (0.1,1.05) -- (0.15,1.35) -- (0.15,1.65);
\draw[line1] (-0.1,-0.1) -- (-0.1,0.9) -- (0.05, 1.75) -- (0.25,1.75) -- (0.25,1.35)
-- (0.1,-0.1) -- (-0.1,-0.1); 
\draw[line1, <-, thick] (-0.1,0.15) -- ++(-0.5,0) node[anchor=east]{polygon surrounding the trajectory}; 
\node[node2] at (0,-0.5) {+};
\node[node2] at (0,-1) {-};
\node[node2] at (0,-1.5) {+};
\end{scope}

\begin{scope}[xshift=0cm, yscale=1, yshift=-1cm]
\foreach \Point in {(0,0), (0.3,0.3), (0.5,0.6), (0.6,0.9), (0.5,1.3), (0.2,1.5), (-0.1, 1.6)}{
    \node[node1] at \Point {\textbullet};
}
\draw[node1] (0,0) -- (0.3,0.3) -- (0.5,0.6) -- (0.6,0.9) -- (0.5,1.3) -- (0.2,1.5) -- (-0.1, 1.6);
\draw[line1] (-0.1,-0.2) -- (0.4,0.3) -- (0.6,0.6) -- (0.7,0.9) -- (0.6,1.4) -- (0.3,1.6) -- (-0.3, 1.8) -- (-0.1,-0.2);
\node[node2] at (0,-0.5) {+};
\node[node2] at (0,-1) {+};
\node[node2] at (0,-1.5) {+};
\end{scope}

\end{tikzpicture}

\end{figure}
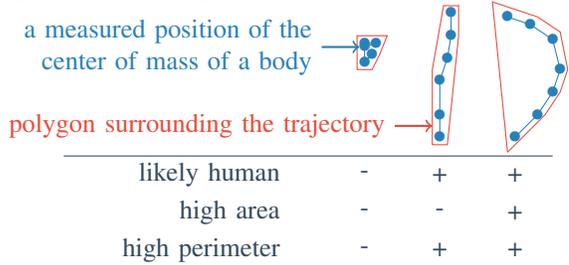

Firstly, an equipment such as a fridge or a shelf is sometimes misidentified as a human body. These false detections are static since these equipments are inanimate. Hence, considering the perimeter of the smallest convex polygon surrounding the points, these false detections can be removed. 
The area bounded by the same polygon could also be used but is less relevant as shown in Fig. \ref{fig:area-vs-perimeter}. When a trajectory is straight, the area can be low whereas the perimeter will be high. Hence the perimeter is used to detect static points characteristic of this first kind of false detections.

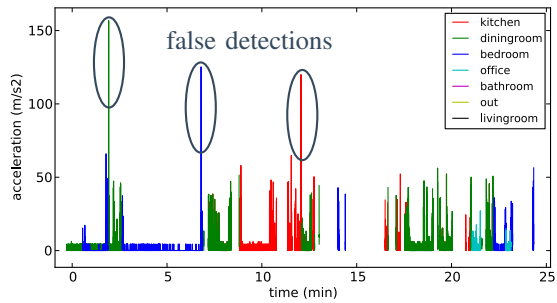
\begin{figure}
\centering
\caption{High accelerations of the center of mass point out false detections}
\label{fig:acceleration}
\begin{tikzpicture}
\tikzstyle{line1}=[black1, thick];
\node[line1] at (0,0) {\includegraphics[width=8.5cm]{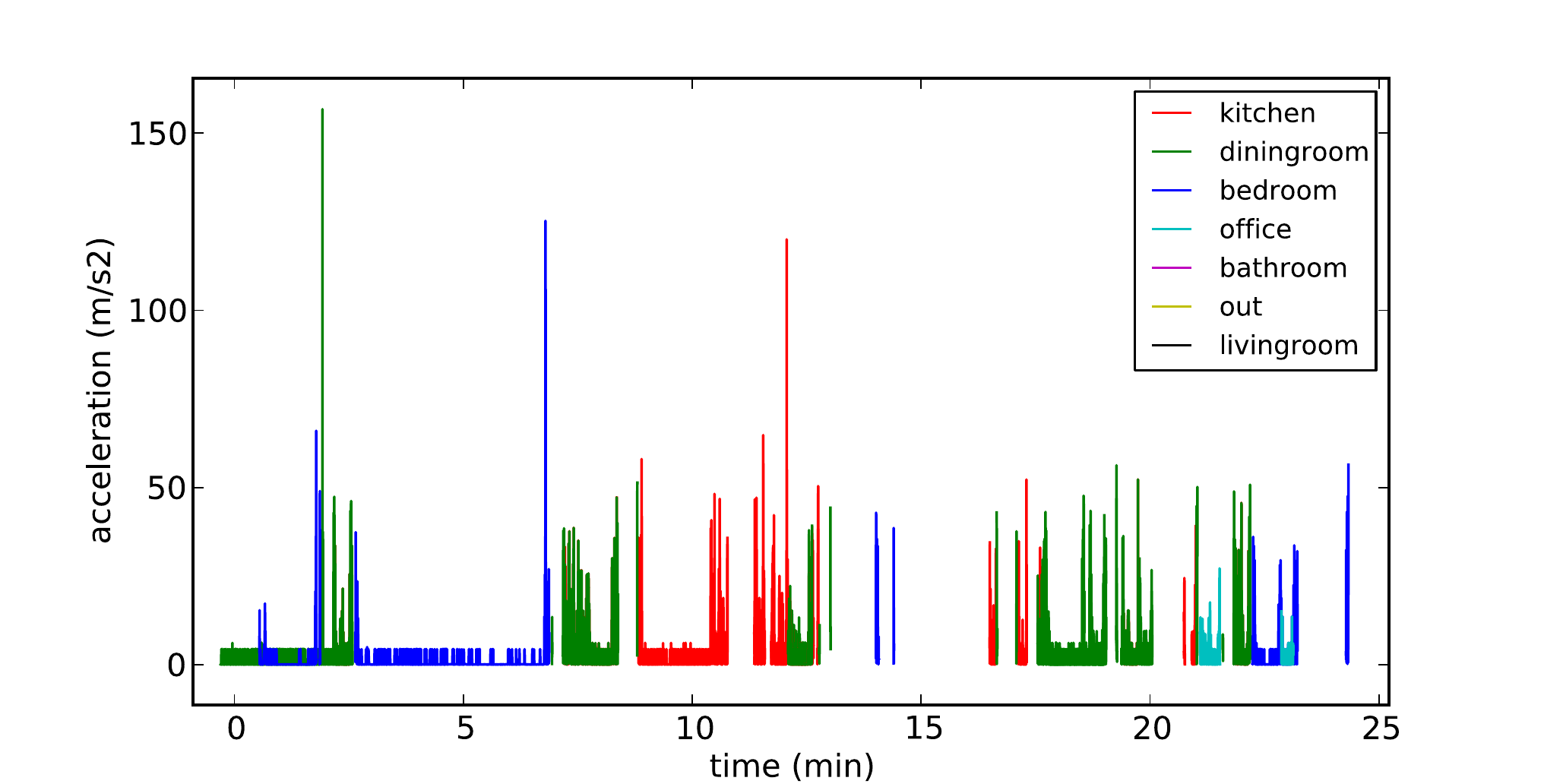}};
\draw[line1] (0.1,0.3) ellipse (0.2cm and 0.6cm);
\draw[line1] (-2.47,1.0) ellipse (0.2cm and 0.6cm);
\draw[line1] (-1.25,0.4) ellipse (0.2cm and 0.6cm);
\node[line1] at (-0.6,1.3) {false detections};
\end{tikzpicture}

\end{figure}

Secondly, when two detected people (possibly fake) are close to each other, the algorithm can mix their trajectories. This error creates a discontinuity in the trajectory which produces a high acceleration of the center of mass of the detected person (Fig. \ref{fig:acceleration}).

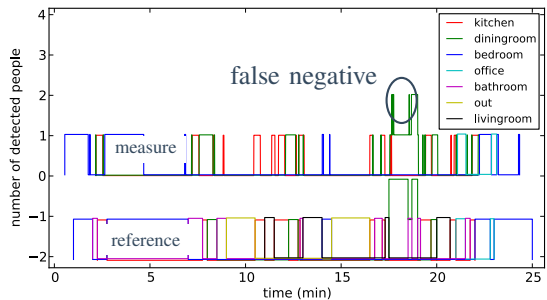
\begin{figure}
\centering
\caption{The filtered ambulatogram (up) and the reference (down)}
\label{fig:filtered-ambulatogram}
\begin{tikzpicture}
\tikzstyle{line1}=[black1, thick];
\tikzstyle{node1}=[line1, font=\scriptsize, fill=white];

\node at (0,0) {\includegraphics[width=8.5cm]{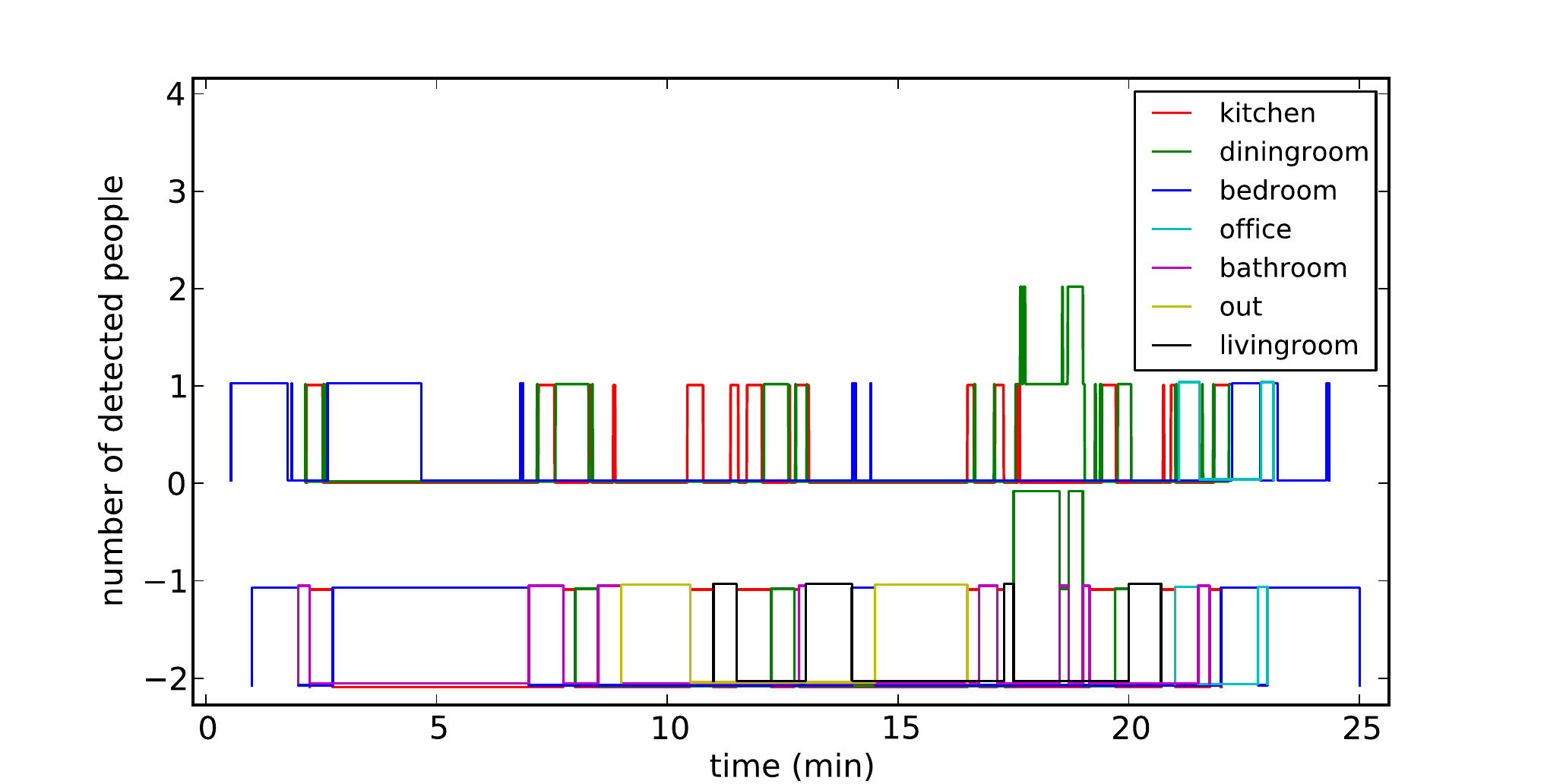}};
\node[node1] at (-1	.9,-0.15) {measure};
\node[node1] at (-1	.9,-1.35) {reference};

\draw[line1] (1.5,0.5) ellipse (0.2cm and 0.3cm);
\node[line1] at (0.2,0.8) {false negative};

\end{tikzpicture}
\end{figure}

By removing the trajectories with high accelerations or only composed of static points, a second ambulatogram is obtained (Fig. \ref{fig:filtered-ambulatogram}). The thresholds for the acceleration (50 m.s$^{-2}$) and for the perimeter (1 m) where set empirically from the experiments. 
The detection report is summarized in table \ref{tab:false-positive-filtered} where the ratio are based on the sequences duration.
All the false detections were removed improving the specificity as expected, but some extra false negative where encountered lowering the sensitivity. 
These were caused by a detection of very short sub-sequences instead of a continuous one. 
Hence, every subsequence covers a very small area (the person is sitting, thus not moving) and the sequence is considered as an artifact and removed.

\begin{table}
\centering
  \caption{Report on filtered data}
  \label{tab:false-positive-filtered}
  \begin{tabular}{| c | c | c |}
	\hline
	 & Sensitivity  & Specificity \\ \hline
	Raw data & 86\%	&	78\% \\ \hline
	Filtered data & 68\%	&	100\%  \\ \hline
  \end{tabular}

\end{table}

\section{Discussion \& Conclusion}

The described system is able to detect the simultaneous presence of several people in the same area. These areas have been defined to infer a reduced set of possible activities. 
Hence, these classifications provide an indicator of simultaneous similar task execution. This one could be a relevant tool for the detection and monitoring of collaborative activity. More scenarios must be played and analysed in order to get statistically representative results.  

Thanks to the processing of the raw detections, the specificity is much better than initially, but the sensitivity could be improved. The low sensitivity is mainly due to the loss of the detection of a static person, especially when lying down on the bed. An analysis of the position before and after a sequence with no detection could give the opportunity to fill this blank.

Finally, this system needs a priori knowledge: a map of the apartment, the position of the Kinects, and the task specific areas. If the system is to be installed in many different apartments, then some of this knowledge should be deduced from a learning phase to reduce the setup duration and the induced errors.

\addtolength{\textheight}{-10cm}   




\section*{ACKNOWLEDGMENT}
The authors would like to thank the Institute of Nanotechnologies of Lyon (INL) for the special grant attributed to Lo\"ic Sevrin for his PhD thesis.


\bibliographystyle{IEEEtran}
\bibliography{../../docear/bibtex/ref2}

\end{document}